\pdfoutput=1

\documentclass[prl,amsmath,amssymb,aps,superscriptaddress,showpacs,
   twocolumn]{revtex4-1}

\usepackage{graphicx}
\usepackage{dcolumn}
\usepackage{bm}
\usepackage[utf8]{inputenc}

\usepackage{siunitx}
\usepackage{braket}

\newcommand{\vsd}{\ensuremath{V_{\text{sd}}}}
\newcommand{\vg}{\ensuremath{V_{\text{gate}}}}

\begin{document}

\title{Universality of the Kondo effect in quantum dots with ferromagnetic 
leads}
\author{M. Gaass} \email{markus.gaass@physik.uni-regensburg.de}
\affiliation{Institute for Exp.~and Applied
Physics, University of Regensburg, 93040 Regensburg, Germany}

\author{A. K. Hüttel}
\affiliation{Institute for Exp.~and Applied
Physics, University of Regensburg, 93040 Regensburg, Germany}

\author{K. Kang}
\affiliation{Institute for Exp.~and Applied
Physics, University of Regensburg, 93040 Regensburg, Germany} 
\affiliation{Department of Physics, Chonnam National University, Gwang-Ju
500-757, Korea}

\author{I. Weymann}
\affiliation{Physics Department, ASC, and CeNS,
Ludwig-Maximilians-Universität, 80333 Munich, Germany}
\affiliation{Department of Physics, Adam Mickiewicz University, 
61-614 Pozna\'n, Poland}

\author{J. von Delft}
\affiliation{Physics Department, ASC, and CeNS,
Ludwig-Maximilians-Universität, 80333 Munich, Germany}

\author{Ch. Strunk}
\affiliation{Institute for Exp.~and Applied
Physics, University of Regensburg, 93040 Regensburg, Germany}

\date{\today}

\begin{abstract} 
  We investigate quantum dots in clean single-wall carbon nanotubes
  with ferromagnetic PdNi-leads in the Kondo regime. In most odd
  Coulomb valleys the Kondo resonance exhibits a pronounced splitting,
  which depends on the tunnel coupling to the leads and an external
  magnetic field $B$, and only weakly on gate voltage. Using numerical
  renormalization group calculations, we demonstrate that all salient
  features of the data can be understood using a simple model for the
  magnetic properties of the leads. The magnetoconductance at zero bias
  and low temperature depends in a universal way on $g \mu_\text{B}
  (B-B_\text{c}) / k_\text{B} T_\text{K}$, where $T_\text{K}$ is the Kondo
  temperature and $B_\text{c}$ the external field compensating the
  splitting.
\end{abstract}

% 73.23.Hk 	Coulomb blockade; single-electron tunneling
% 73.63.Fg 	Nanotubes (electronic transport)
% 72.15.Qm 	Scattering mechanisms and Kondo effect (METALS AND ALLOYS)
% 72.25.-b 	Spin polarized transport

\maketitle

The Kondo effect resulting from the exchange interaction of a single spin
with a bath of conduction electrons \cite{Hewson1993}, is one of the
archetypical phenomena of many-body physics. Its competition with
ferromagnetism and possible applications in spintronics \cite{Fabian2004}
have raised wide interest in the past few years. The Kondo effect in
quantum dots \cite{GoldhaberGordon1998PRL,Kouwenhoven1998} has, in recent
experiments, been investigated in the presence of ferromagnetic (FM) leads
\cite{Pasupathy2004a,Hauptmann2008,Hofstetter2010}. It was found that the
Kondo resonance, usually observed at zero bias in the odd Coulomb blockade
(CB) valleys, is split into two peaks at finite bias \cite{Pasupathy2004a}.
The splitting consists of a term depending logarithmically on gate voltage
\cite{Hauptmann2008,Hofstetter2010}, and, as demonstrated here, a second
term nearly independent of gate voltage. These phenomena were predicted
theoretically \cite{Martinek2003,Choi2004,Martinek2005a,Sindel2007},
attributing the splitting of the Kondo resonance to a tunneling induced
exchange field, which results from the magnetic polarization of the leads. 
So far no detailed and quantitative comparison of the measured conductance
with the theory has been undertaken to verify whether the simplistic
description of FM leads used in
Refs.~\onlinecite{Martinek2003,Choi2004,Martinek2005a,Sindel2007} has
quantitative predictive power. The latter would be needed for future
spintronics applications that exploit the lead-induced local spin
splitting, e.g., spin filtering.

In this Letter we present low-temperature transport measurements of a
single wall carbon nanotube quantum dot with PdNi leads. We concentrate on
the less studied gate-independent contribution of the exchange splitting of
the Kondo resonance and attribute it to the saturation magnetization of the
contact material. We show that the evolution of the conductance with
magnetic field and gate voltage can be understood within a simple model for
the magnetization and polarization in the FM leads, by presenting numerical
renormalization group (NRG) calculations for this model, using parameters
extracted from experiment. Moreover, by comparing resonances of different
transparency, we demonstrate a universal scaling of the magnetic field
dependence of the Kondo conductance, proving that the magnetization of the
leads can indeed be viewed as an exchange field, which acts analogously to
an external magnetic field.

\emph{Experimental setup.---} The nanotubes are grown by chemical vapor
deposition on a highly doped silicon substrate with a \SI{200}{\nano\meter}
thermally grown oxide layer at its surface \cite{nature-kong:878}. The
contact electrodes with a thickness of roughly \SI{45}{\nano\meter} are
subsequently structured by electron beam lithography and consist of
$\text{Pd}_{0.3}\text{Ni}_{0.7}$, known to generate highly transparent
contacts \cite{Sahoo2005}. Transport measurements were done in a dilution
refrigerator with a base temperature of \SI{25}{\milli\kelvin}. The current
$I$ is measured in a two point geometry with a voltage bias \vsd\ applied
to the source contact.

\begin{figure}
\includegraphics[width=\columnwidth]{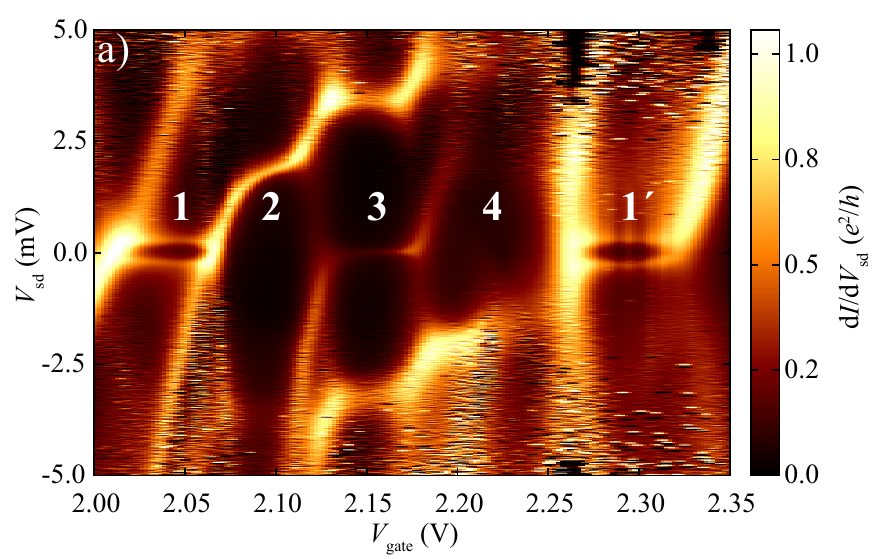}\vspace{-2mm}
\caption{
(Color online) Differential conductance $\text{d}I/\text{d}\vsd$ versus
bias voltage \vsd\ and gate voltage \vg\ at $T= \SI{25}{\milli\kelvin}$ and
$B = \SI{0}{\tesla}$. The CB regions are numbered for future reference.
\vspace{-5mm}}
\label{Fig:ChargingDiagram}
\end{figure}
The differential conductance $G=\text{d}I/\text{d}\vsd$ is plotted in
Fig.~\ref{Fig:ChargingDiagram} in color scale, providing the typical
charging diagram of a quantum dot \cite{kouwenhoven}. Our device
exhibits regular CB oscillations and a clear fourfold symmetry
characteristic for carbon nanotubes \cite{prl-oreg:365}.  Coupling to
the contacts is strong, leading to broad resonance lines and a variety
of higher-order processes. Fig.~\ref{Fig:ChargingDiagram} implies the
following parameters for the nanotube quantum dot
\cite{prl-oreg:365,Sapmaz2005}: a charging energy $U \simeq
\SI{5}{\milli\electronvolt}$, a level separation $\Delta \simeq
\SI{9.5}{\milli\electronvolt}$, and a subband mismatch of about $\delta
\simeq \SI{1}{\milli\electronvolt}$. The tunnel coupling $\Gamma$ between
leads and quantum dot can be inferred from the line width of the
conductance peak. Between valleys 1 and 2 in Fig.~\ref{Fig:ChargingDiagram}
we obtain a full width at half maximum (FWHM) of $\Gamma=
\SI{1.1}{\milli\electronvolt} \gg k_\text{B} T$.

A striking feature visible in Fig.~\ref{Fig:ChargingDiagram} are lines of
enhanced conductance at small, approximately constant bias values in every
second CB diamond. We attribute these lines to a spin-1/2 Kondo conductance
anomaly, split into two peaks at small opposite bias values due to the
presence of FM contacts. The peak distance $2 \Delta\varepsilon$ at the
center of the diamond can be related to a magnetic field scale
$B=\Delta\varepsilon/g\mu_{\rm B}$ via the Zeeman effect. For valley
1 in Fig.~\ref{Fig:ChargingDiagram} this gives approximately
\SI{2}{\tesla} for $g=2$. Some resonances with very low conductance
exhibit no measurable splitting.

\begin{figure}
\includegraphics[width=\columnwidth]{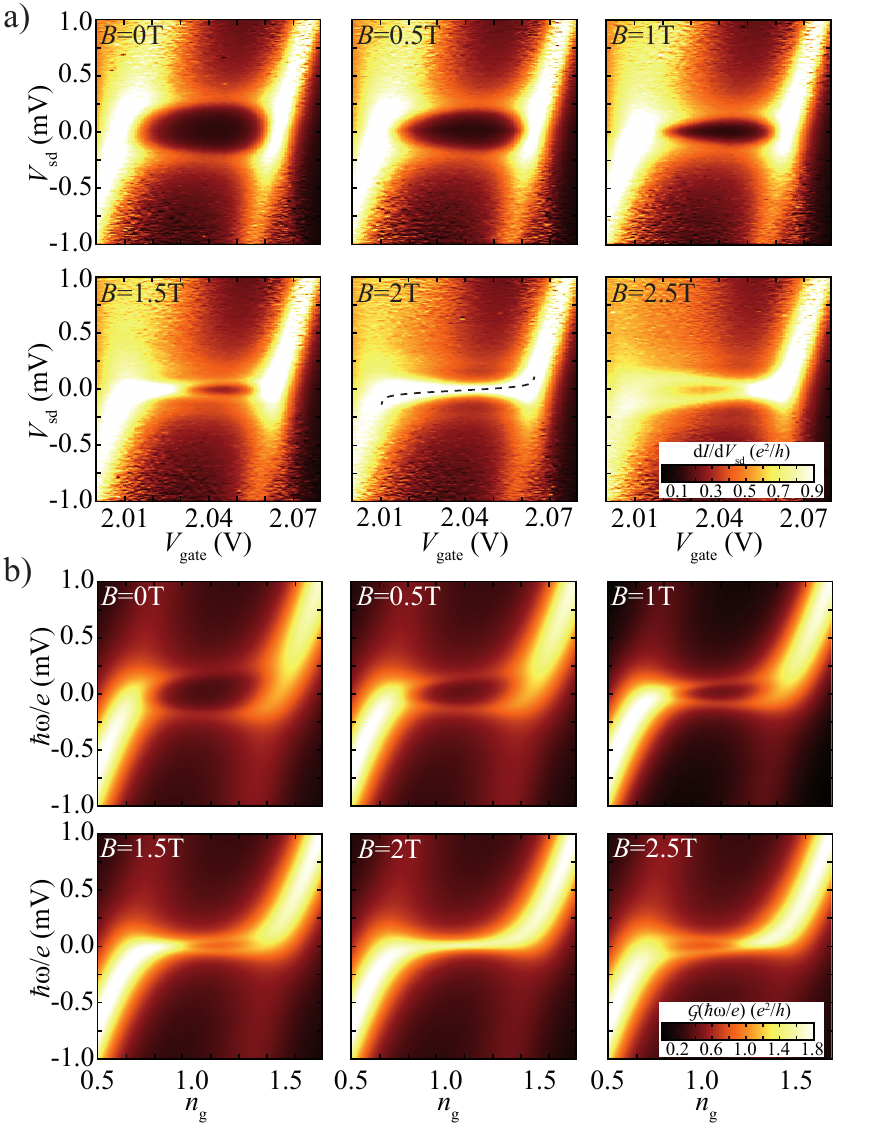} \vspace{-6mm}
\caption{ (Color online) Differential conductance versus
    source-drain and gate voltage, in CB region 1 of
    Fig.~\ref{Fig:ChargingDiagram}, for six magnetic field values. We
    compare here (a) experimental data  with (b) NRG results for the
    normalized zero-temperature spectral function $\mathcal{G}
    =\sum_\sigma \pi \Gamma_\sigma(0) A_\sigma(\omega)$, obtained
    using the model and parameters described in the
    text. $n_\text{g} = 1/2 - \varepsilon_\text{d}/U$ is the dimensionless
    gate potential. The dashed line in the panel for $B =
    \SI{2}{\tesla}$ in (a) indicates the gate-dependent contribution
    from the polarization for ${\cal P}=10\%$ (see
    text). \vspace{-5mm} }
\label{Fig:CompTheoryExper}
\end{figure}

\emph{Main features of $B$-dependence:---}
Figure~\ref{Fig:CompTheoryExper}(a) displays detailed measurements of
valley 1 from Fig.~\ref{Fig:ChargingDiagram} for different values of
external magnetic fields almost parallel to the nanotube axis. The main
observations are: (i) The dominant contribution to the splitting is
independent of \vg. (ii) As the field strength increases, the splitting is
reduced (observed in all investigated cases) until only a single apparent
peak remains. This field value is referred to as compensation field (here
$B_{\rm c}\simeq\SI{2}{\tesla}$), since the dominant gate-independent part
of the splitting is compensated. At higher fields the peak splits again. 
(iii) Despite (i), we observe nevertheless a slight gate dependence, in
particular near $B_{\rm c}$. This is most clearly reflected by the fact
that the touching point of the two resonances moves from the left side of
the CB diamond for $B<B_{\rm c}$ (cf. $B = \SI{1.5}{\tesla}$) to the right
side for $B>B_{\rm c}$ (cf. $B = \SI{2.5}{\tesla}$). (iv) The gate
dependence increases in strength very close to the edges of the CB diamond.

\emph{Tunneling-induced level shifts.---} The presence of the splitting,
its dependence on magnetic field, and a potential gate dependence can be
consistently explained by the renormalization of the quantum dot
level energy due to charge fluctuations between the dot and the leads.
Since the density of states (DOS) in the FM contacts is spin dependent, the
same is true for the charge fluctuations and the corresponding energy
renormalization. Hence, the renormalization results in a spin splitting of
the dot level $\Delta\varepsilon \equiv \delta\varepsilon_{\uparrow} -
\delta\varepsilon_{\downarrow} - g\mu_B B$. For a single impurity Anderson
model, the correction $\delta \varepsilon_{\sigma}$ ($\sigma
=\,\uparrow,\downarrow$) from second order perturbation theory is (see,
e.g., Ref.~\cite{Martinek2005a}):
\begin{equation}
\label{Eq:EnergyCorrection}
\delta\varepsilon_{\sigma} \simeq -\frac{1}{\pi} \int
d\omega\left(\frac{\Gamma_{\sigma}(\omega)[1-f(\omega)]}{\omega-
\varepsilon_{\text{d},\sigma}}+\frac{\Gamma_{-\sigma}(\omega)f(\omega)}{
\varepsilon_{\text{d},-\sigma}+U-\omega} \right).
\end{equation}
Here, $\varepsilon_{\text{d},\sigma}=\varepsilon_\text{d}\mp
g\mu_\text{B}B/2$ is the quantum level energy for spin $\sigma$, $U$ the
charging energy, and $\Gamma_\sigma(\omega)$ the spin-dependent tunneling
rate. $f(\omega)$ is the Fermi function.

From Eq.~(\ref{Eq:EnergyCorrection}) one sees that the splitting is not 
only a consequence of properties at the Fermi surface, but of the full DOS.
The first and second terms in Eq.~(\ref{Eq:EnergyCorrection}) describe
electron- or hole-like processes, meaning fluctuations between the states
$\ket{1,\sigma}$ and $\ket{2}$ or $\ket{0}$, respectively (the numeral
denotes the charge occupation of the quantum dot). The spin dependent
energy corrections $\delta\varepsilon_{\sigma}$ are negative, as always in
second order perturbation theory. Consequently, the spin direction that
favors fluctuations more strongly will have lower renormalized energy.

\begin{figure}
\includegraphics[width=8cm]{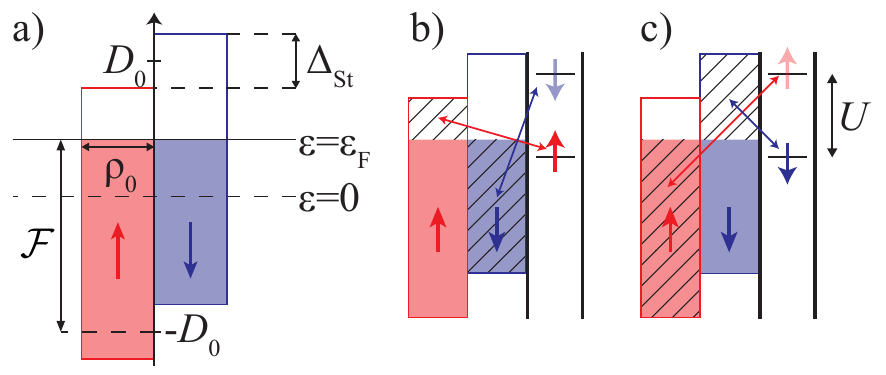} \vspace{-3mm}
\caption{(Color online) Schematic of the level renormalization process. (a)
We assume flat bands $\rho_{\sigma}(\omega) = \rho_{0}$ with bandwidth
$D_0$, shifted with respect to each other by a constant
Stoner splitting $\Delta_{\rm St}$. The filling fraction ${\cal F}$
determines the Fermi energy $\varepsilon_\text{F}$. (b) Charge fluctuations
for a spin up electron on the dot, involving the empty states of the spin
up band and the occupied states of the spin down band (hatched areas). (c)
Analogous situation for a spin down electron on the dot; the available 
phase space (hatched) is larger than in (b). \vspace{-3mm}}
\label{Fig:Schematic}
\end{figure}

\emph{Effect of Magnetization.---} First we assume a shift between bands of
equal and constant DOS for different spin directions, $\rho_{\uparrow} =
\rho_{\downarrow}=\rho_0$, described by a constant Stoner splitting
$\Delta_{\rm St}$, see Fig.~\ref{Fig:Schematic}. The tunneling induced
splitting $\Delta\varepsilon^{\mathcal{(M)}}$ due to $\Delta_{\rm St}$ is
directly related to the saturation magnetization $\mathcal{M} \equiv
(n_{\uparrow}-n_{\downarrow})/N_a = \Delta_{\rm St}/(2D_0)$. Here
$n_{\sigma}=\rho_0(D_0+\varepsilon_{\rm F}\pm\Delta_{\rm St}/2)$ is the
number of spin-$\sigma$ electrons, $\varepsilon_{\rm F}$ the Fermi energy,
and $N_a$ the number of states per atom and spin. Starting from
Eq.~(\ref{Eq:EnergyCorrection}) we can deduce the spin orientation of the
dot ground state as follows: Figs.~\ref{Fig:Schematic}(b) and (c) show the
phase space available (hatched) for quantum charge fluctuations for a spin
up or a spin down electron residing on the quantum dot, respectively.
Comparing the total sizes of hatched areas for (b) and (c), the phase
space is larger for the latter, thus favoring spin down. Therefore, $\Delta
\varepsilon^{({\cal M})}$ is always negative, meaning that the energy
correction due to a finite Stoner splitting is larger for the minority
spin, i.e. the ground state spin is oriented opposite to the magnetization
of the leads. This explains why the splitting is always initially reduced
(never increased) when an external field is applied (cf. observation (ii)
listed above; the magnetization direction follows that of the field in our
setup). The size of this effect depends on the Stoner splitting,
i.e., on $\mathcal{M}$. For $|\varepsilon_{\rm d}|, \varepsilon_{\rm d}+U
\ll D_0, \Delta_{\rm St}$, which is compatible with the experiment, we
obtain, in extension of \cite{Sindel2007},
\begin{equation}
\label{Eq:Magnetization} \Delta\varepsilon^{(\mathcal{M})} \simeq
\frac{\Gamma}{2\pi} \ln \left[\frac{(1-\mathcal{M})^2-(2{\cal
F}-1)^2}{(1+\mathcal{M})^2-(2{\cal F}-1)^2}\right],
\end{equation}
where ${\cal F}=(1+\varepsilon_{\rm F}/D_0)/2$ is the filling fraction of
the band. This shift is independent of the gate voltage (explaining
observation (i)) because the position $\varepsilon_\text{d}$ of the
level is very close to $\varepsilon_\text{F}$, while the integration in
Eq.~(\ref{Eq:EnergyCorrection}) ranges over a large fraction of the d-band.

SQUID measurements allow us to determine a magnetic moment of $\mu =
0.58\,\mu_{\rm B}/\text{atom}$ for our PdNi alloy \cite{steininger},
implying an effective magnetization of $\mathcal{M}=0.116$. Ab-initio
calculations of the band structure provide a filling fraction of ${\cal F} = 0.853$
\cite{mankovsky}. Thus, we estimate $\Delta\varepsilon^{(\mathcal{M})}
\simeq -175\,\mu{\rm eV}$. In transport spectroscopy this would lead to
conductance peaks split at zero external field by $2 \Delta\varepsilon^{\rm
theo}/e = 2 |
\Delta\varepsilon^{(\mathcal{M})}/e | \simeq350\,\mu{\rm V}$. Given the
simplicity of our model, this agrees reasonably well with the
experimentally determined peak distance of $2 \Delta\varepsilon^{\rm exp}/e
\simeq550\,\mu{\rm V}$. For a more weakly coupled resonance the predicted
peak distance of $60\,\mu{\rm V}$ agrees similarly with the experimentally
found value of $105\,\mu{\rm V}$.

\emph{Effect of polarization.---} The case of $\rho_\uparrow \neq
\rho_\downarrow$, implying nonzero polarization ${\cal P} = (\rho_\uparrow
- \rho_\downarrow)/(\rho_\uparrow+\rho_\downarrow)$ at
$\varepsilon_\text{F}$, has already been discussed in
Refs.~\cite{Sindel2007,Hauptmann2008} and earlier publications referenced
there. Assuming a flat band with a spin-dependent DOS, e.g.\
$\rho_{\uparrow}>\rho_{\downarrow}$ but zero $\mathcal{M}$ for simplicity,
quantum charge fluctuations renormalize the quantum dot level
depending on its position relative to $\varepsilon_\text{F}$. This
contribution shows a logarithmic divergence for $\varepsilon_{\rm d}
\rightarrow 0$ and $\varepsilon_{\rm d} + U \rightarrow 0$
\cite{Hauptmann2008, Sindel2007}, resulting in the up- and downward bending
of the compensated conductance peak towards the corners of the diamond (cf.
observation (iv)).

\emph{Numerical results.---} The quality of our model is reflected by
the close correspondence of Figs.~\ref{Fig:CompTheoryExper}(a) and
(b).  Fig.~\ref{Fig:CompTheoryExper}(b) presents high-quality NRG
results for the spectral function $A(\omega)$ \footnote{Though
  $A(\omega)$ is calculated for an equilibrium model at zero
  temperature, its $\omega$-dependence mimics (qualitatively, if not
  quantitatively) the $V_{\rm sd}$-dependence of the conductance for a
  two-lead setup.} versus the normalized gate voltage $n_{\rm g}$,
calculated for a single-lead Anderson model with the DOS shown in
Fig.~3 but with $\rho_\uparrow \neq \rho_\downarrow$, using {\em full}
density-matrix NRG \cite{Weichselbaum,Legeza}. Using the measured
parameters of the quantum dot and modeling the ferromagnetism in the
leads by taking ${\cal P} = 10 \%$ and ${\cal M} = 0.116$, very good
agreement with experiment is found, except for the background current
at high \vsd, which results probably from cotunneling processes
involving higher levels not included into the model. In the experiment
as well as in the numerical data, at $B= \SI{2}{\tesla}$ the
gate-independent contribution, cf. Eq.~(\ref{Eq:Magnetization}), is
fully compensated and the crossing point of the resonances lies in the
center of the CB diamond. Here, only the (weak) gate-dependent
contribution from ${\cal P}$ remains (cf. observation
(iii)), indicated in Fig.~\ref{Fig:CompTheoryExper} ($B=\SI{2}{\tesla}$) by
a dashed line. By varying ${\cal P}$ and comparing the shape of the
conductance peak at $B_{\rm c}$ between experiment and theory, we
estimate ${\cal P} \simeq 10\%$.

\begin{figure}
\includegraphics[width=\columnwidth]{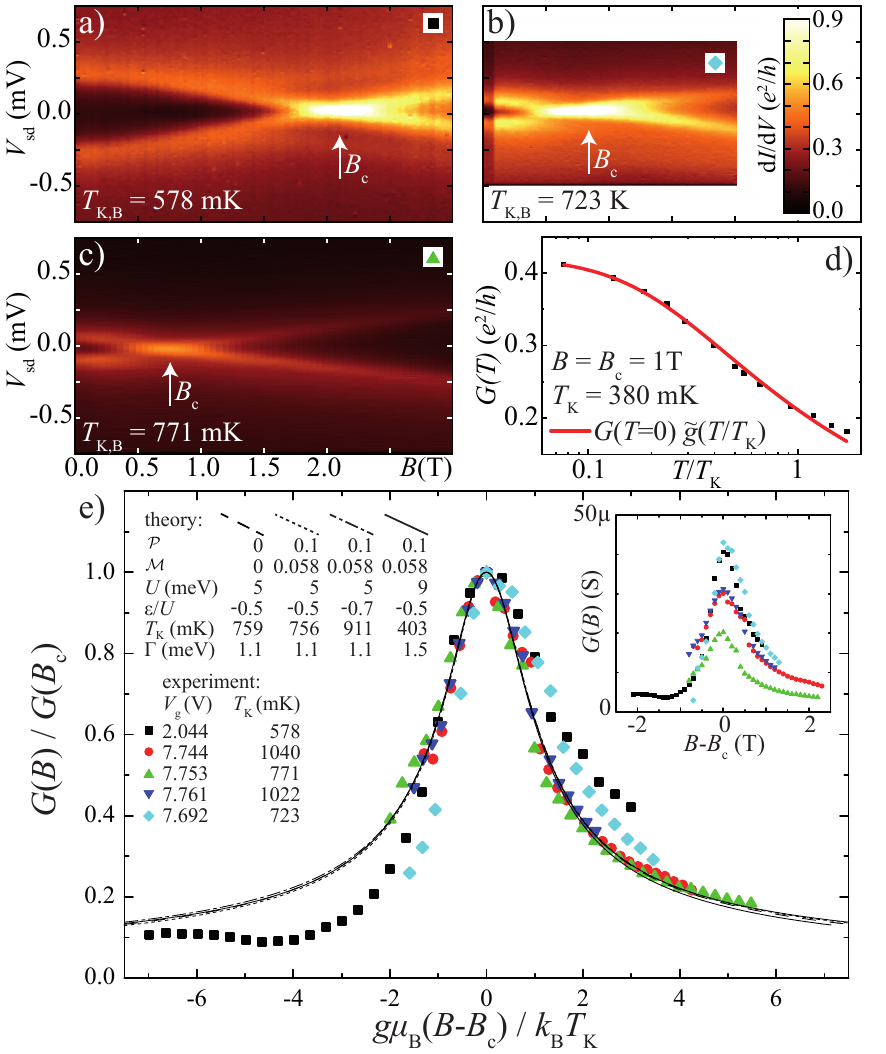} \vspace{-6mm}
\caption{(Color online) (a)-(c) Magnetic field dependence of the
  splitting for different charge states. Vertical arrows denote the
  compensation field $B_\text{c}$. (d) Scaling plot of $G(T)$ vs.\
  $T/T_\text{K}$ (symbols), at $B=B_\text{c}$, for the charge state shown
  in Fig.~\ref{Fig:Universality}(c); the solid line
  gives $G(0)$ times the universal function $\tilde{g}(T/T_{\rm
  K})$ discussed in the main text. (e) Scaled zero bias
  conductance $G(B) / G(B_\text{c})$ plotted against the effective
  normalized field $\delta \tilde B$, at fixed $T = 50\,\text{mK} \ll
  T_\text{K}$ ($100\,\text{mK}$ for $\blacksquare$).  
 Lines represent
  NRG calculations for several parameter sets. Inset: $G(B)$ vs.\
  $(B-B_\text{c})$ before scaling. \vspace{-6mm}
\label{Fig:Universality}}
\end{figure}

\emph{Universality.---} For a quantum dot coupled to normal leads, the
normalized zero-bias conductance is a universal function (1) of $\tilde T =
T/T_\text{K}$ at zero field, and (2) of $\tilde B = g \mu_\text{B} B/
k_\text{B} T_\text{K}$ at zero temperature. (We define the Kondo
temperature $T_\text{K}$ via $G(T=T_{\rm K})/G(0)=1/2$ at zero field). We
find, quite remarkably, that both these universal features are recovered
also for ferromagnetic leads, if $B$ is replaced by the effective field
$\delta B = B-B_\text{c}$. Regarding (1), Fig.~\ref{Fig:Universality}(d)
shows that at the compensation field,  $B=B_\text{c}$, the temperature
dependence of the normalized conductance, $G(T)/G(0)$, agrees with the
often-used semi-empirical formula $\tilde g (\tilde T)= [1 + (2^{1/s}-1)
\tilde T^2]^{-s}$, with $s\simeq0.22$ \cite{GoldhaberGordon1998PRL}.
Although the latter behavior is well-established for dots coupled to normal
leads, its emergence here is nontrivial: it demonstrates that despite the
ferromagnetic environment, local spin symmetry can indeed be fully
restored by fine-tuning the field to $B_\text{c}$.

The magnetic field dependence (2) has so far attracted much less attention
\cite{Quay2007}. To explore it, Fig.~\ref{Fig:Universality}(a)-(c) shows
$G(B)$ at fixed $T \ll T_\text{K}$ for several charge states differing in
$T_\text{K}$, $B_\text{c}$, and $\Delta\varepsilon(B=0)$. The position of
the conductance peak roughly follows the Zeeman law, with slight deviations
in the vicinity of the compensation field $B_\text{c}$ \cite{Quay2007}. We
find that $B_\text{c}$ and $T_\text{K}$ vary independently for different
charge states, implying different couplings $\Gamma$.

According to (2), $G(B)/G(B_\text{c})$ should be a universal function of
$\delta \tilde B = g\mu_\text{B}(B-B_\text{c})/ k_\text{B}T_\text{K}$. 
The lines in Fig.~\ref{Fig:Universality}(e) show this curve, calculated by
NRG for four different sets of model parameters, yielding a good scaling
collapse. Symbols show experimental data for $G(B)/G(B_\text{c})$ vs.\
$\delta \tilde B$, for several different gate voltages, with $T_\text{K}$
extracted by numerical
fitting to the NRG results. For three data sets taken
on the same charge state (circles and triangles), scaling works very well
and agreement with theory is excellent; for the other two sets (squares,
diamonds), the quality of scaling is reduced at higher $\delta\tilde B$ by
an asymmetric background contribution to the magnetoconductance.
Nevertheless, the overall agreement between theory and experiment
shows that the model correctly captures the universal, sample-independent
features of $G(B)$ as function of $\delta \tilde B$.

\emph{Conclusions.---} We have performed a quantitative comparison of
the conductance of quantum dots with FM contacts, in the Kondo regime,
with model NRG calculations. The quantitative agreement between
experimental and numerical data lends strong support to the scenario
proposed in Refs.~\onlinecite{Martinek2003,Sindel2007}: the exchange
field induced by magnetic contacts causes the local level to be split
by an amount $\Delta \varepsilon$, which adds a constant offset to
the Zeeman splitting induced by an external magnetic field. When this
offset is compensated by a suitably fine-tuned field $B_\text{c}$,
universal scaling features of the Kondo effect are recovered.  The
control of $\Delta \varepsilon(B)$ through the selection of materials with
proper $\cal{M}$ and $\cal{P}$ (high or small $\cal{M}$ optimize peak
splitting or gate tunability, respectively) may prove useful for future
applications in spintronics. 

We acknowledge fruitful discussions with M.\ Grifoni, S.\ Koller, J.\
Paaske, M.~R.\ Wegewijs, and H.~S.~J.\ van der Zant, and thank S.\
Mankovsky for providing results of ab-initio calculations. This project has
been supported by the DFG within SFB~689 and the NIM excellence cluster.
I.\ W.\ was supported by the Polish MSHE and the Humboldt Foundation, and
K.\ K. by the NRF of Korea under Grant No. 2009-0072595 and by the LG
Yeonam Foundation. In the final stage of the preparation of this manuscript
we became aware of a similar study in quantum dots with nonmagnetic
contacts \cite{dgg}.

\vspace{-6mm}

\bibliography{kondopaper}

\end{document}